\documentclass[journal]{IEEEtran}
\IEEEoverridecommandlockouts
\usepackage{lineno}
\usepackage{hyperref}
\usepackage{cite}
\usepackage{amsmath,amssymb,amsfonts,cases} 
\usepackage{amsmath}
\usepackage{amsthm}

\usepackage{graphicx}
\usepackage{textcomp}
\usepackage{xcolor}
\usepackage{graphicx}
\usepackage{float}
\usepackage{subfigure}
\usepackage{amsmath,mleftright}
\usepackage{amsfonts,amssymb}
\usepackage{mathrsfs}
\usepackage{mathtools}
\usepackage{algorithm}
\usepackage{algorithmic}
\usepackage{bm}
\usepackage{multirow}
\usepackage{array}
\usepackage{amssymb}
\usepackage{amsmath}
\usepackage{cite}
\usepackage{url}
\usepackage{xcolor}
\usepackage{cite,graphicx,amsmath,amssymb}
\usepackage{subfigure}
\usepackage{fancyhdr}
\usepackage{mdwmath}
\usepackage{mdwtab}
\usepackage{caption}
\usepackage{amsthm}
\usepackage{setspace}
\usepackage{bm}
\usepackage{mathtools}
\usepackage{dsfont}
\usepackage{bbm}
\usepackage{framed}

\newtheorem{theorem}{Theorem}

\newtheorem{lemma}{Lemma}

\newtheorem{corollary}{Corollary}

\allowdisplaybreaks[4]
\makeatletter
\newcommand{\biggg}{\bBigg@{3}}
\newcommand{\Biggg}{\bBigg@{3.5}}
\makeatother
\makeatletter
\renewcommand{\maketag@@@}[1]{\hbox{\m@th\normalsize\normalfont#1}}%
\makeatother
\def\BibTeX{{\rm B\kern-.05em{\sc i\kern-.025em b}\kern-.08em
    T\kern-.1667em\lower.7ex\hbox{E}\kern-.125emX}}
    \expandafter\def\expandafter\normalsize\expandafter{%
    \normalsize%
    \setlength\abovedisplayskip{4pt}%
    \setlength\belowdisplayskip{4pt}%
    \setlength\abovedisplayshortskip{2pt}%
    \setlength\belowdisplayshortskip{2pt}%
}
\begin{document}
\title{Beamforming Gain with Single-RF Movable Arrays}
\author{Zhenqiao Cheng, Chongjun Ouyang, Hao Jiang, Xingqi Zhang, and Arumugam Nallanathan\vspace{-10pt}
\thanks{Z. Cheng is with the 6G Research Centre, China Telecom Beijing Research Institute, Beijing 102209, China (e-mail: zhenqiao.cheng@engineer.com).}
\thanks{C. Ouyang, H. Jiang, and A. Nallanathan are with the School of Electronic Engineering and Computer Science, Queen Mary University of London, London, E1 4NS, U.K. (e-mail: \{c.ouyang, hao.jiang, a.nallanathan\}@qmul.ac.uk).}
\thanks{X. Zhang is with the Department of Electrical and Computer Engineering, University of Alberta, Edmonton AB, T6G 2R3, Canada (e-mail: xingqi.zhang@ualberta.ca).}}
\maketitle

\begin{abstract}
A single-radio-frequency (RF) movable array is investigated, in which all movable elements are driven by a single RF chain with equal amplitude and equal phase. The achievable beamforming gain enabled by antenna placement is analyzed. Linear beamforming gain scaling with the number of antennas is shown to be achievable in single-path channels, while coherent-combining conditions and aperture requirements are established for multipath channels. For multiuser transmission, the optimal max-min power allocation is derived in closed form, based on which an element-wise coordinate-search algorithm is developed for antenna placement design. Numerical results validate the analysis and reveal a fundamental tradeoff: beamforming gains can be achieved through antenna placement alone, but only at the expense of increased aperture resources.
\end{abstract}


\section{Introduction}
Movable antennas exploit local spatial channel variations through controlled antenna displacement \cite{wong2020fluid,new2024tutorial,zhu2024movable,zhu2025tutorial}. This introduces an additional electromagnetic-domain degree of freedom beyond conventional array processing and aligns with the emerging tri-hybrid view of multi-antenna architectures, where digital, analog, and reconfigurable-antenna operations jointly shape the transmitted field \cite{heath2026trihybrid}. Existing movable-array designs typically employ either one radio-frequency (RF) chain per antenna element \cite{zhu2023movable,cheng2024sum,cheng2024enabling} or phase shifters (PSs) \cite{xiong2025secure,zhang2026hybrid} after a power splitter, and jointly optimize beamforming weights and antenna positions.

\begin{figure}[!t]
\centering
\includegraphics[height=0.12\textwidth]{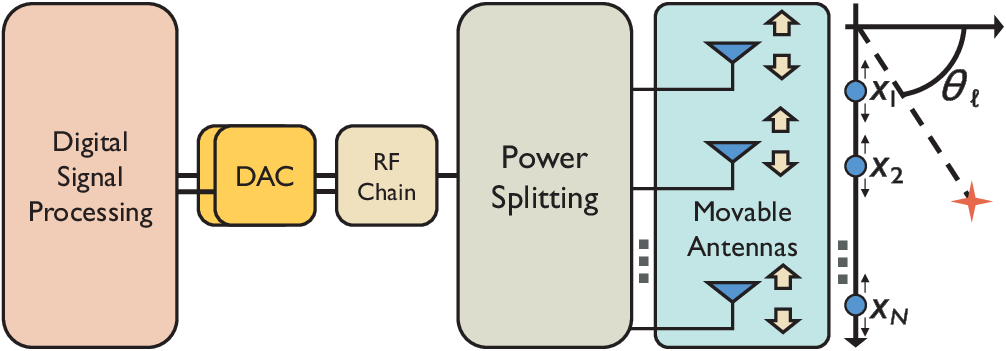}
\caption{Illustration of a single-RF movable array.}
\label{Figure0}
\vspace{-15pt}
\end{figure}

This article considers a more restrictive architecture in which all movable elements are driven by a single RF chain through a power splitter with equal amplitude and equal phase, as illustrated in {\figurename} {\ref{Figure0}}. Consequently, antenna placement becomes the sole beamforming mechanism. At first sight, the achievable gain appears limited because arbitrary per-element phase control is unavailable \cite{heath2026trihybrid,heath2016overview}. The central question is therefore \emph{whether meaningful beamforming gains can be achieved through antenna placement alone, and if so, under what conditions}. As shown in this article, coherent combining can indeed be realized through appropriate antenna placement. The resulting gain, however, is fundamentally constrained by aperture availability. 

To answer this question, a compact theory for the beamforming gain of single-RF movable arrays is developed. For single-path channels, we show that the achievable gain scales linearly with the number of antennas, while the required aperture scales linearly as well. For two-path channels, we show that exact coherent combining requires commensurate spatial frequencies, whereas irrational frequency ratios can only be approximated at an aperture cost characterized by Dirichlet approximation \cite{cassels1957diophantine}. For general multipath channels, we show that strongest-path alignment induces a Dirichlet-kernel response and leads to a spatial-frequency filtering effect. We further derive an aperture-availability law that quantifies the aperture required to support a prescribed fraction of propagation directions. Finally, for time-division multiple access (TDMA) multiuser transmission, we obtain the optimal max-min power allocation in closed form and solve the resulting antenna-placement problem. Overall, the analysis reveals a fundamental tradeoff: \emph{the beamforming gains offered by phase-control circuitry can be partially recovered through flexible antenna placement, but the price is aperture}.

\section{System Model}
Consider a transmitter equipped with a single RF chain and $N$ movable antenna elements deployed over a one-dimensional aperture $[0,A]$, as shown in {\figurename} {\ref{Figure0}}. The position of element $n$ is denoted by $x_n$, and the placement vector is $\mathbf{x}=[x_1,\ldots,x_N]^{\mathsf{T}}$. To avoid severe mutual coupling, the antenna positions satisfy
\begin{align}
|x_n-x_{n'}|\ge d_{\min}, \qquad n\neq n',
\end{align}
where $d_{\min}=\lambda/2$ and $\lambda$ denotes the wavelength \cite{ivrlavc2010toward}. The RF signal is equally divided among the $N$ elements through a power splitter. Since only one RF chain is available, beamforming is achieved solely through antenna placement.

Let $\mathcal{K}\triangleq \{1,\ldots,K\}$ denote the user set served by the movable array via TDMA. For user $k\in\mathcal{K}$, the spatial channel field at location $x\in[0,A]$ is modeled as follows \cite{ma2023compressed}:
\begin{align}
h_k(x)=\sum_{\ell=1}^{L_k}\alpha_{k,\ell}{\rm{e}}^{{\rm{j}}\kappa_{k,\ell}x},
\label{eq_channel}
\end{align}
where $L_k$ is the number of resolvable paths, $\alpha_{k,\ell}$ is the complex path gain, and $\kappa_{k,\ell}\triangleq
\frac{2\pi}{\lambda}
\sin(\theta_{k,\ell})$ is the spatial frequency associated with the directional angle $\theta_{k,\ell}\in[-\frac{\pi}{2},\frac{\pi}{2}]$.

The channel sampled by the movable array is $\mathbf h_k(\mathbf x)\triangleq[h_k(x_1),\ldots,h_k(x_N)]^{\mathsf{T}}$. When user $k$ is scheduled with power $P_k$, its signal-to-noise ratio (SNR) is given by $\gamma_k(\mathbf x,P_k)\triangleq P_kG_k(\mathbf x)$, where
\begin{align}
G_k(\mathbf x)=\frac{|\mathbf 1^{\mathsf{T}}\mathbf h_k(\mathbf x)|^2}{N\sigma^2}=\frac{1}{N\sigma^2}\left|\sum_{n=1}^{N}h_k(x_n)\right|^2,
\label{snr}
\end{align}
and $\sigma^2$ is the noise power. The factor $\frac{1}{N}$ accounts for the equal power split across the antenna elements. Since the same RF waveform reaches every element, $\mathbf 1^{\mathsf T}\mathbf h_k(\mathbf x)$ is the only coherent-combining term available to the transmitter. Therefore, all beamforming degrees of freedom are embedded in the antenna placement vector $\mathbf x$.
\section{Single-User Gain Laws}\label{Section: Single-User Gain Laws}
We first study one scheduled user and drop the user index. Define the achievable beamforming gain as follows:
\begin{align}\label{gain}
{\mathsf{G}}\triangleq\frac{1}{N}\left\lvert \sum_{n=1}^{N}h_k(x_n)\right\rvert^2=\frac{1}{N}\left\lvert \sum_{\ell=1}^{L}\alpha_{\ell}\sum_{n=1}^{N}{\rm{e}}^{{\rm{j}}\kappa_{\ell}x_n}\right\rvert^2.
\end{align}
The antenna placement problem is formulated as follows:
\begin{align}\label{Problem_Single_User}
\max_{\mathbf{x}}~{\mathsf{G}},\quad {\rm{s.t.}}~ \lvert x_n-x_{n'}\rvert\geq d_{\min},~x_n\in[0,A].
\end{align}
The structure of \eqref{gain} depends critically on the number of propagation paths. In the sequel, we respectively investigate the single-path, two-path, and general multipath cases to characterize the achievable beamforming gain and the associated aperture requirements.
\subsection{Single-Path Channel}
We begin with the single-path case, i.e., $L=1$. The beamforming gain reduces to the following:
\begin{align}
{\mathsf G}
=
\frac{|\alpha_1|^2}{N}
\left|
\sum_{n=1}^{N}
{\rm e}^{{\rm j}\kappa_1 x_n}
\right|^2.
\label{single_path_gain}
\end{align}

The maximum gain is attained when all sampled channel coefficients have identical phases. This condition is satisfied by the uniform placement as follows:
\begin{align}
x_n
=
x_1
+
(n-1)d,
\qquad
d
=
\frac{2\pi p}{|\kappa_1|},
\quad
p\in{\mathbbmss Z}^{+},
\label{single_path_placement}
\end{align}
provided that $x_N\le A$. Under \eqref{single_path_placement}, all antenna elements observe the same channel phase. Therefore,
\begin{align}
{\mathsf G}
=N|\alpha_1|^2.
\label{single_path_scaling}
\end{align}
Equation \eqref{single_path_scaling} shows that a single-RF movable array achieves the same linear array gain as coherent phase shifting. 

The linear gain scaling comes at the cost of aperture. The required spacing is given by
\begin{align}
d
=
\frac{2\pi}{|\kappa_1|}
=
\frac{\lambda}{|\sin\theta_1|}
\geq
\lambda>
d_{\min},
\label{single_path_spacing}
\end{align}
which automatically satisfies the minimum-spacing constraint. Consequently, the aperture required for coherent combining is
\begin{align}
A_{\mathsf{req}}
= (N-1)d
=
\frac{(N-1)\lambda}{|\sin\theta_1|}.
\label{single_path_aperture}
\end{align}

Equation \eqref{single_path_aperture} reveals that the aperture requirement grows linearly with the number of antenna elements. Moreover, when the propagation direction approaches broadside, i.e., $|\sin\theta_1|\rightarrow 0$, the required aperture increases rapidly. Therefore, the achievable beamforming gain and the required aperture are fundamentally coupled in single-RF movable arrays. The result also provides an alternative interpretation of beamforming. In conventional phased arrays, PSs electronically compensate the propagation phases across antenna elements. In contrast, the proposed architecture achieves the same compensation geometrically through antenna placement. As a result, beamforming capability is obtained through aperture resources rather than dedicated phase-control circuitry.
\subsection{Two-Path Channel}
We next consider the two-path channel
\begin{align}
h(x)
=
\alpha_1{\rm e}^{{\rm j}\kappa_1x}
+
\alpha_2{\rm e}^{{\rm j}\kappa_2x}.
\end{align}
The beamforming gain satisfies
\begin{align}
{\mathsf G}
\le
N(|\alpha_1|+|\alpha_2|)^2,
\label{twopath_upper}
\end{align}
where equality requires all antenna elements to observe the same path-combined phase. Unlike the single-path case, phase alignment must now be achieved simultaneously for two distinct spatial frequencies. To see when this is possible, consider a uniform placement with inter-antenna spacing $d$. Simultaneous phase alignment requires 
\begin{align}
\lvert\kappa_1\rvert d
=
2\pi p,
\qquad
\lvert\kappa_2\rvert d
=
2\pi q,
\label{common_period}
\end{align}
for some integers $p,q\in{\mathbbmss Z}^{+}$. Therefore,
\begin{align}
\frac{ \lvert\kappa_1\rvert}{\lvert\kappa_2\rvert}=\frac{p}{q}\in{\mathbbmss{Q}},
\label{rational_condition}
\end{align}
which implies that the two spatial frequencies must share a common spatial period. This is the key difference from the single-path case, where only one spatial period needs to be matched. 

When \eqref{rational_condition} holds, one feasible spacing is
\begin{align}
d
= \frac{2\pi p}{\lvert \kappa_1\rvert}=
\frac{2\pi q}{\lvert \kappa_2\rvert}
\geq
\lambda\min\{p,q\}>
d_{\min}.
\end{align}
If the aperture satisfies $(N-1)d\le A$, all antenna elements can observe identical path-combined phases and the upper bound in \eqref{twopath_upper} becomes achievable.

When $\frac{ \lvert\kappa_1\rvert}{\lvert\kappa_2\rvert}$ is irrational, exact simultaneous alignment is impossible over any nonzero spacing. Nevertheless, irrational ratios can be approximated arbitrarily well by rational numbers. In particular, Dirichlet's approximation theorem guarantees infinitely many integer pairs $(p,q)$ satisfying \cite{cassels1957diophantine}
\begin{align}
\left|
\frac{ \lvert\kappa_1\rvert}{\lvert\kappa_2\rvert}-
\frac{p}{q}
\right|
<
\frac{1}{q^2}.
\end{align}
Choosing $d
=
\frac{2\pi q}{\lvert\kappa_2\rvert}$ aligns the second path exactly and yields a residual phase mismatch on the first path bounded by
\begin{align}
\left|
\lvert \kappa_1\rvert d-
2\pi p
\right|
<
\frac{2\pi}{q}.
\label{residual_phase}
\end{align}

Equation \eqref{residual_phase} reveals a fundamental gain-aperture tradeoff. Increasing $q$ reduces the residual phase mismatch and brings the beamforming gain closer to the coherent upper bound. However, the required spacing also increases proportionally to $q$. Consequently, the aperture requirement becomes
\begin{align}
A_{\mathsf{req}}
=
(N-1)\frac{2\pi q}{\lvert\kappa_2\rvert}.
\end{align}

To maintain near-coherent combining across $N$ antenna elements, the accumulated phase mismatch across the array must remain small. Since the residual error in \eqref{residual_phase} scales as ${\mathcal O}(\frac{1}{q})$, the aggregate mismatch $N\left|\lvert \kappa_1\rvert d-2\pi p\right|$ scales approximately as $\mathcal O(\frac{N}{q})$. Therefore, near-full coherent gain requires $q$ to increase at least proportionally to $N$. Substituting this scaling into the aperture expression yields $A_{\mathsf{req}}=\mathcal O(N^2)$.

Therefore, unlike the single-path case, adding more antenna elements alone does not guarantee coherent gain. The aperture must increase simultaneously to support the spatial resolution required for aligning multiple propagation paths. In particular, while the single-path channel requires an aperture that scales linearly with $N$, the two-path channel may require a quadratic aperture scaling to approach full coherent combining.

\subsection{Multipath Channel}
For a general multipath channel with $L\geq 3$, exact simultaneous alignment becomes increasingly restrictive. 
\subsubsection{Strongest-Path Alignment}
Motivated by the preceding analysis, we consider a constructive design that aligns the strongest path and use the resulting beamforming gain as an achievable benchmark. Without loss of generality, assume $|\alpha_1|
=
\max_{\ell=1,\ldots,L}
|\alpha_\ell|$, and choose the uniform placement $d=\frac{2\pi}{|\kappa_1|}$. The resulting beamforming gain is
\begin{align}
{\mathsf{G}}&=\frac{1}{N}\left\lvert N\alpha_{1}{\rm{e}}^{{\rm{j}}\kappa_{1}x_1}+\sum_{\ell=2}^{L}\alpha_{\ell}{\rm{e}}^{{\rm{j}}\kappa_{\ell}x_1}\sum_{n=1}^{N}{\rm{e}}^{{\rm{j}}\kappa_{\ell}(n-1)d}\right\rvert^2\\
&=\frac{1}{N}\left\lvert N\alpha_{1}+\sum_{\ell=2}^{L}\alpha_{\ell}{\rm{e}}^{{\rm{j}}\Delta_{\ell}x_1}\sum_{n=1}^{N}{\rm{e}}^{{\rm{j}}\Delta_{\ell}(n-1)d}\right\rvert^2\\
&=N\lvert\alpha_{1}\rvert^2\left\lvert 1+\sum_{\ell=2}^{L}\frac{\alpha_{\ell}}{\alpha_1}{\rm{e}}^{{\rm{j}}\Delta_{\ell}x_1}D_N(\Delta_{\ell}d)\right\rvert^2,\label{dirichlet_gain}
\end{align}
where $\Delta_{\ell}\triangleq\kappa_{\ell}-\kappa_{1}$ is the spatial-frequency offset, and
\begin{align}
D_N(u)
\triangleq
\frac{1}{N}
\sum_{n=1}^{N}
{\rm{e}}^{{\rm{j}}(n-1)u}
=
{\rm{e}}^{-{\rm{j}}\frac{(N-1)u}{2}}
\frac{\sin(\frac{Nu}{2})}
{N\sin(\frac{u}{2})}
\end{align}
is the normalized Dirichlet kernel.

Equation \eqref{dirichlet_gain} separates the coherent contribution of the strongest path from the leakage generated by the remaining paths. In a dominant Rician channel satisfying $\frac{|\alpha_\ell|}{|\alpha_1|}\ll1$ for $\ell\ne1$, the leakage term becomes negligible and ${\mathsf G}
\approx
N|\alpha_1|^2$. Therefore, the strongest-path-aligned placement preserves the linear gain scaling derived for the single-path channel.

More generally, the contribution of path $\ell$ is weighted by the Dirichlet kernel $D_N(\Delta_\ell d)$. As $N\rightarrow\infty$, $D_N(\Delta_\ell d)\rightarrow0$ unless $\Delta_{\ell}d=2\pi q$ for $q\in{\mathbbmss{Z}}$. Thus, only paths satisfying the strongest-path alignment condition contribute coherently. 

For finite $N$, define $\delta_\ell\triangleq(\Delta_\ell d)\bmod 2\pi$. For $\delta_\ell\approx0$, $|D_N(\delta_\ell)|
\approx
\left|
{\rm sinc}
\left(
\frac{N\delta_\ell}{2}
\right)
\right|$. A path contributes at least an $\eta$ fraction of its coherent value only if $|D_N(\delta_\ell)|
\geq
\eta$, which asymptotically requires $|\delta_\ell|
\lesssim
\frac{2c_\eta}{N}$, where $|{\rm sinc}(c_\eta)|=\eta$.
Therefore, the strongest-path-aligned movable array behaves as a spatial-frequency filter whose effective bandwidth shrinks proportionally to $\frac{1}{N}$. Only paths belonging to the same spatial-frequency cluster as the dominant path contribute coherently, whereas paths outside this cluster become asymptotically resolvable and contribute negligibly to the beamforming gain.
\subsubsection{Aperture Availability}
The preceding analysis reveals that beamforming gain is obtained by trading aperture resources for phase-control circuitry. For the strongest-path-aligned construction, the required aperture is $A_{\mathsf {req}}
=
\frac{(N-1)\lambda}
     {|\sin\theta_1|}$; see \eqref{single_path_aperture}. Assume that the dominant-path direction $\theta_1$ is uniformly distributed over $[-\frac{\pi}{2},\frac{\pi}{2}]$. Then the aperture required to support a fraction $p\in[0,1]$ of dominant-path directions is
\begin{align}
A_{p}=
\frac{(N-1)\lambda}
{\sin(\frac{\pi }{2}(1-p))}
=
\frac{(N-1)\lambda}
{\cos\left(\frac{\pi p}{2}\right)}.
\end{align}

\begin{figure}[t]
\centering
\includegraphics[width=0.4\textwidth]{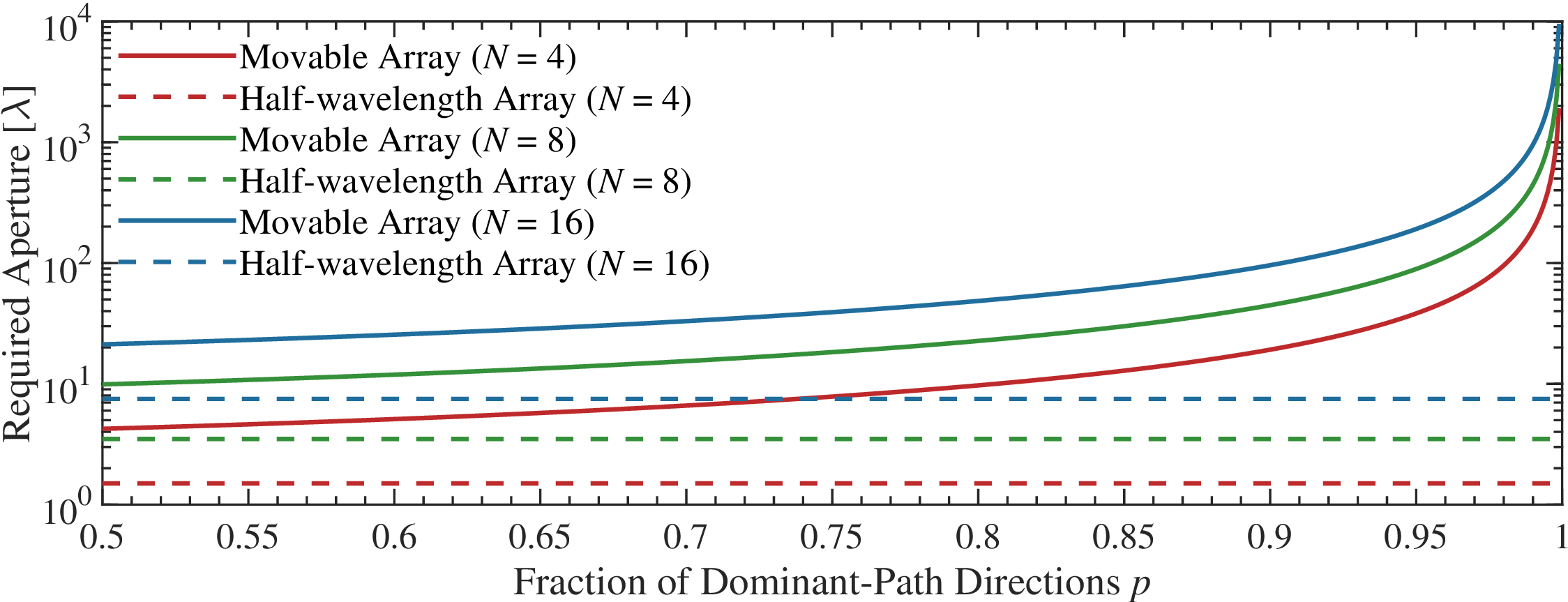}
\caption{Required aperture for strongest-path alignment.}
\label{fig:Aperture_Single_User}
\vspace{-15pt}
\end{figure}

{\figurename} {\ref{fig:Aperture_Single_User}} illustrates $A_{p}$ versus the coverage fraction $p$. The required aperture increases rapidly as $p$ approaches one. For example, supporting 95\% and 99\% of dominant-path directions requires approximately $12.7(N-1)\lambda$ and $63.7(N-1)\lambda$, respectively. These values are substantially larger than those of a conventional half-wavelength-spaced array, whose aperture scales only as $0.5(N-1)\lambda$. Therefore, the linear beamforming gain achieved by strongest-path alignment comes at the expense of a significantly enlarged aperture. This observation highlights a fundamental tradeoff between beamforming gain and aperture resources in single-RF movable arrays. It also explains the performance gap between aperture-limited and aperture-rich deployments observed in the numerical results.

\subsubsection{Element-Wise Placement Optimization}
If the antenna placement is not restricted to the strongest-path-aligned construction, problem \eqref{Problem_Single_User} can be solved numerically via an element-wise coordinate-search algorithm. At each iteration, $x_n$ is optimized while keeping the remaining positions fixed. The resulting subproblem is given by
\begin{align}
\max_{x_n}~&{\mathsf G}^{(n)}(x_n)
\triangleq
\frac{1}{N}\left\lvert \sum_{\ell=1}^{L}\alpha_{\ell}{\rm{e}}^{{\rm{j}}\kappa_{\ell}x_n}+\sum_{n'\ne n}\sum_{\ell=1}^{L}\alpha_{\ell}{\rm{e}}^{{\rm{j}}\kappa_{\ell}x_{n'}}\right\rvert^2\nonumber\\
{\rm s.t.}~&x_n\in[0,A],\ |x_n-x_{n'}|\geq d_{\min},~n\neq n'.\nonumber
\end{align}
We next adopt a one-dimensional grid search over the feasible interval. Specifically, the interval $[0,A]$ is discretized into a uniform $Q$-point grid ${\mathcal Q}\triangleq\{
0,\frac{A}{Q-1},\ldots,A\}$. To enforce the minimum-spacing constraint, define $\hat{{\mathcal Q}}_n\triangleq\{x|x\in{\mathcal Q},|x-x_{n'}|<d_{\min},n'\neq n\}$ as the set of invalid points. The updated antenna position is then chosen as follows:
\begin{align}
x_n^\star\triangleq\arg\max\nolimits_{x_n\in{\mathcal Q}\setminus\hat{{\mathcal Q}}_n}{\mathsf G}^{(n)}(x_n).
\end{align}

The antenna positions are updated sequentially until convergence. Since each update cannot decrease the objective and ${\mathsf G}$ is upper bounded, the algorithm converges to a stationary point. The overall complexity scales as ${\mathcal O}(I_{\mathsf {iter}}NQ)$, where $I_{\mathsf {iter}}$ denotes the number of outer iterations. The detailed procedure is summarized in Algorithm \ref{Algorithm1}.

\begin{algorithm}[!t]
\algsetup{linenosize=\tiny}
  \scriptsize
\caption{Element-wise algorithm for solving \eqref{Problem_Single_User}}
\label{Algorithm1}
\begin{algorithmic}[1]
\STATE Initialize ${\mathbf{x}}$
\REPEAT
\FOR{$n\in\{1,\ldots,N\}$}
\STATE Update $x_n$ via one-dimensional search
\ENDFOR
\UNTIL{convergence}
\end{algorithmic}
\end{algorithm}

\section{Multiuser TDMA Design}
We next extend the analysis to a TDMA system. The antenna placement vector $\mathbf x$ is shared by all users, whereas the transmit power may vary across time slots. Unlike the single-user case, a common placement cannot generally align the strongest path of every user simultaneously. To ensure fairness, we consider the max-min SNR problem as follows:
\begin{subequations}
\begin{align}
\max_{{\mathbf{x}}, \{P_k\}} \quad
& \min_{k \in \mathcal{K}}~ \gamma_k({\mathbf{x}}, P_k)=P_kG_k(\mathbf x) \\
\rm{s.t.} \quad
& \sum_{k=1}^{K} P_k \le P,~P_k \ge 0,\label{Power_Cons}\\
& \lvert x_n-x_{n'}\rvert\geq d_{\min},~x_n\in[0,A],\label{Phase_Shifter_Cons}
\end{align}
\end{subequations}
where $P$ denotes the power budget. Introducing an auxiliary variable $t$ yields the equivalent formulation as follows:
\begin{align}\label{Problem_Equal}
\min_{{\mathbf{x}}, \{P_k\}, t} \quad
 -t,\quad 
{\rm{s.t.}} ~
 P_kG_k(\mathbf x) \ge t, ~\eqref{Power_Cons},~\eqref{Phase_Shifter_Cons}.
\end{align}

For a given placement $\mathbf x$, a standard Karush-Kuhn-Tucker (KKT) analysis \cite{boyd2004convex} shows that all users achieve the same SNR at optimum, i.e.,
\begin{align}
\gamma_k({\mathbf{x}}, P_k)=P_kG_k(\mathbf x)=
t,
\qquad
k\in\mathcal K.
\end{align}
Since the power budget is fully utilized, the optimal power allocation is $P_k^\star=\frac{P/G_k(\mathbf x)}{\sum_{j=1}^{K}1/G_j(\mathbf x)}$, and the corresponding max-min SNR is $t^\star(\mathbf x)
=
\frac{P}
{\sum_{k=1}^{K}1/G_k(\mathbf x)}$.

Therefore, the joint power-allocation and antenna-placement problem reduces to the following:
\begin{subequations}\label{multi_place}
\begin{align}
\min_{\mathbf x}~&\sum_{k=1}^{K}\frac{1}{G_k(\mathbf x)}
=\sum_{k=1}^{K}\frac{N\sigma^2}{|\mathbf 1^T\mathbf h_k(\mathbf x)|^2}\\
{\rm{s.t.}}~&\lvert x_n-x_{n'}\rvert\geq d_{\min},~x_n\in[0,A].
\end{align}
\end{subequations}
Problem \eqref{multi_place} can be solved using the element-wise coordinate-search framework developed in the previous section by replacing the objective ${\mathsf G}$ with $\sum_{k=1}^{K}\frac{1}{G_k(\mathbf x)}$. Since each accepted update decreases the objective and the objective is nonnegative, the resulting iterations converge in objective value.

\section{Numerical Results}
\begin{figure}[!t]
    \centering
    \subfigure[${\mathsf{G}}$ vs. $N$. $A=20\lambda$.]
    {
        \includegraphics[height=0.17\textwidth]{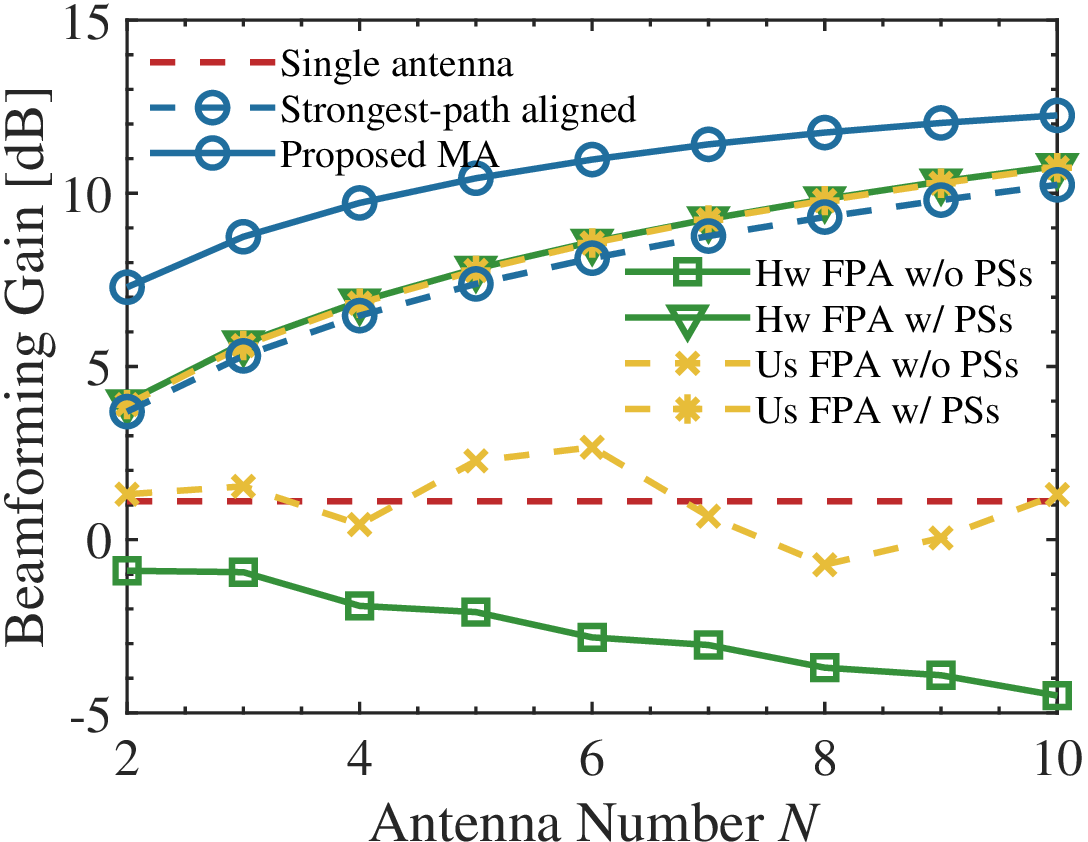}
	   \label{fig1a}	
    }
   \subfigure[${\mathsf{G}}$ vs. $A$. $N=7$.]
    {
        \includegraphics[height=0.17\textwidth]{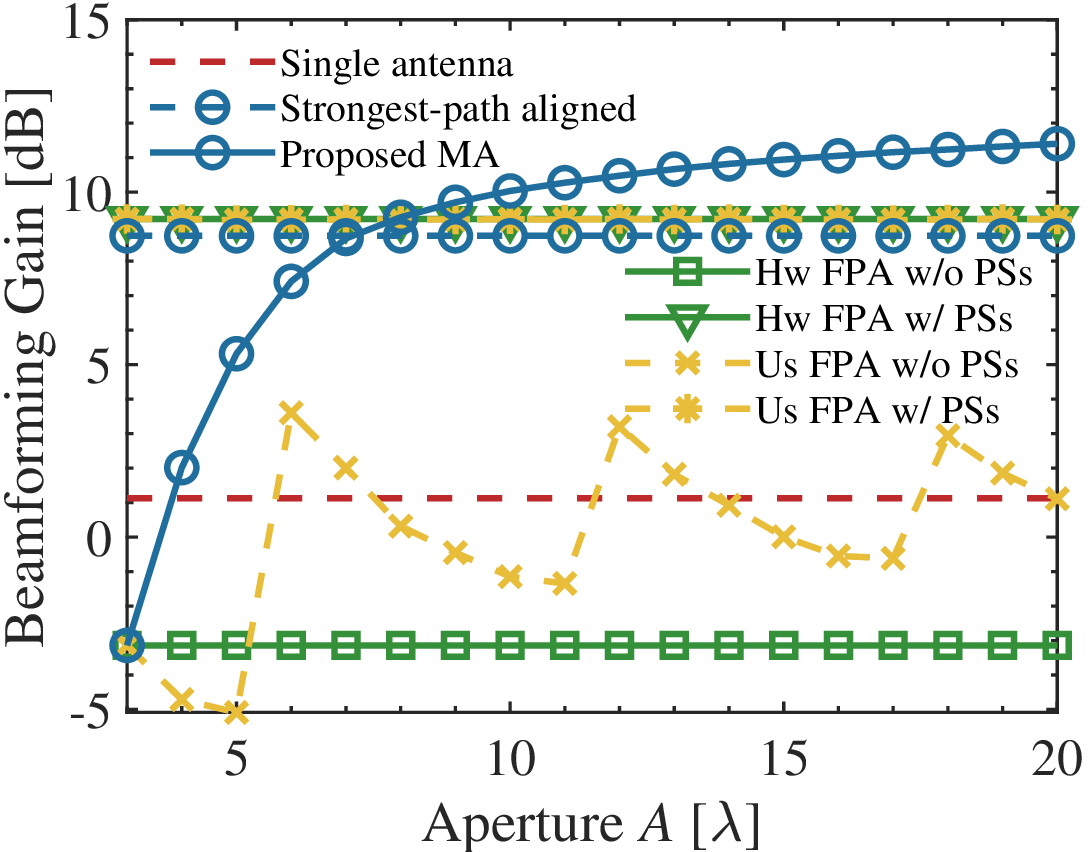}
	   \label{fig1b}	
    }
\caption{Single-user beamforming gain vs. (a) the antenna number and (b) the aperture size.}
    \label{figure1}
    \vspace{-10pt}
\end{figure}

\begin{figure}[!t]
    \centering
    \subfigure[min SNR vs. $N$. $A=20\lambda$.]
    {
        \includegraphics[height=0.17\textwidth]{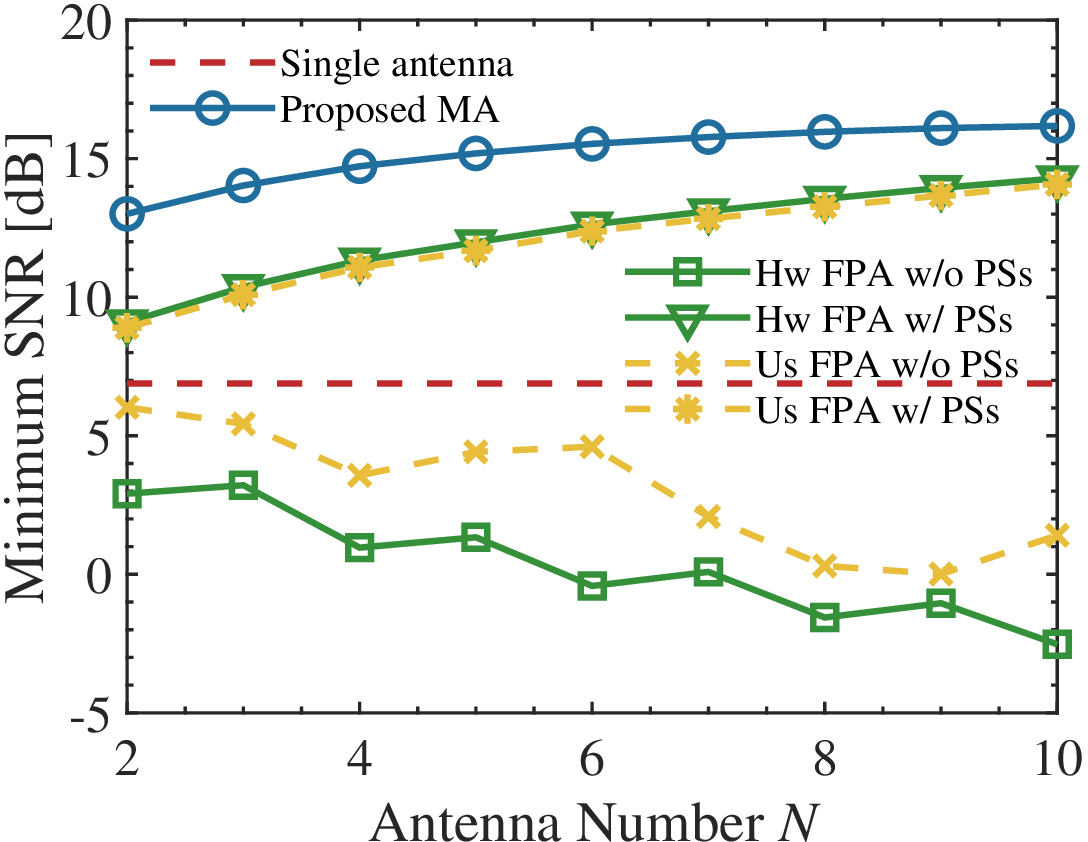}
	   \label{fig2a}	
    }
   \subfigure[min SNR vs. $A$. $N=7$.]
    {
        \includegraphics[height=0.17\textwidth]{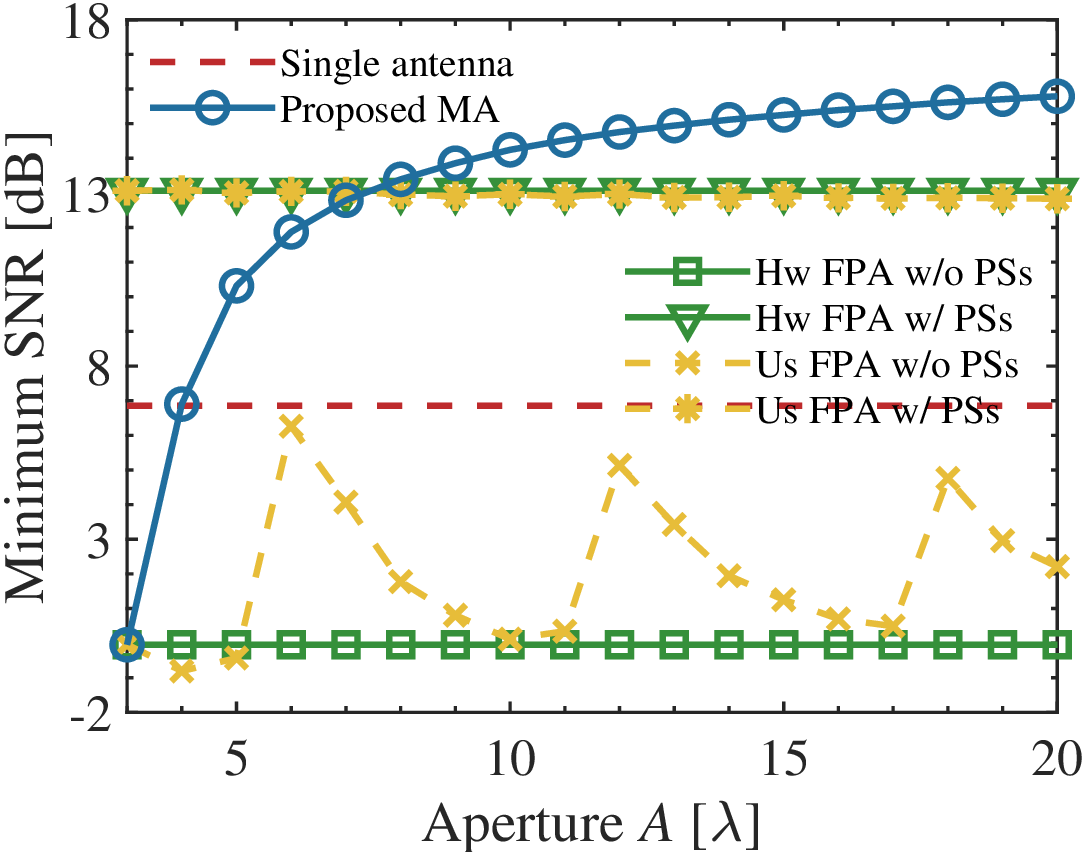}
	   \label{fig2b}	
    }
\caption{Multiuser minimum SNR vs. (a) the antenna number and (b) the aperture size. $K=4$.}
    \label{figure2}
    \vspace{-10pt}
\end{figure}

This section validates the derived results. Unless otherwise specified, each user channel consists of one LoS path and three NLoS paths. The path gains follow $\alpha_{k,1}\sim{\mathcal{CN}}(0,1)$ and $\alpha_{k,\ell}\sim{\mathcal{CN}}(0,10^{-1})$ for $\ell>1$, while the directional angles are independently drawn from the uniform distribution over $[-\frac{\pi}{2},\frac{\pi}{2}]$. The transmit SNR is set to $\frac{P}{\sigma^2}=15$ dB and $Q=10^3$. We compare the proposed MA design with four fixed-array benchmarks: half-wavelength (Hw) FPA without PSs, Hw FPA with PSs, and uniformly-spaced (Us) FPA occupying the same aperture as the MA with and without PSs. The phase shifts are optimized through the element-wise approach.

{\figurename} {\ref{fig1a}} shows the beamforming gain versus the number of antennas for $A=20\lambda$. The beamforming gain achieved by the proposed MA increases monotonically with $N$, which agrees with the analytical results in Section \ref{Section: Single-User Gain Laws}. The strongest-path-aligned construction exhibits similar scaling, confirming that coherent combining can be achieved through antenna placement alone. Since this benchmark assumes sufficient aperture to realize the required spacing, it serves as an achievable reference rather than a fixed-aperture design. Compared with fixed arrays without PSs, the proposed MA achieves substantial performance gains and approaches the phased-array benchmark despite employing only a single RF chain. {\figurename} {\ref{fig1b}} depicts the beamforming gain versus aperture size for a fixed number of antennas. The MA performance improves significantly as the aperture increases. In contrast, the phased-array benchmark is largely insensitive to aperture, whereas the fixed arrays without PSs benefit only marginally from aperture expansion. These results confirm that flexible antenna placement can compensate for the absence of phase-control circuitry, but at the expense of increased aperture requirements.

{\figurename} {\ref{fig2a}} and {\figurename} {\ref{fig2b}} show the minimum user SNR versus the number of antennas and aperture size, respectively. Similar behavior is observed in the multiuser setting. The proposed MA significantly outperforms the fixed-array benchmarks and approaches the phased-array benchmark. The performance gap becomes more pronounced as the available aperture increases.

\section{Conclusion}
This letter analyzed the beamforming gain of single-RF movable arrays. Coherent combining was shown to be achievable through antenna placement alone, and the corresponding gain laws and multiuser design were established. The analysis revealed a fundamental tradeoff: beamforming gains can be recovered through flexible antenna placement, but the price is aperture. In particular, substantially larger apertures may be required than in conventional half-wavelength arrays to achieve comparable beamforming gains.

\clearpage
\newpage
\bibliographystyle{IEEEtran}
\bibliography{mybib_SPL}
\end{document}